\documentclass[%
 longbibliography,
reprint,
showpacs,
 amsmath,amssymb,
 aps,
pre,
showkeys
]{revtex4-1}

\usepackage{graphicx}
\usepackage{dcolumn}
\usepackage{bm}

\begin{document}

\title{Applying allometric scaling to predator-prey systems}

\author{Andreas Eilersen}
  \email{andreaseilersen@nbi.ku.dk}
 \affiliation{Niels Bohr Institute, University of Copenhagen, Blegdamsvej 17, 2100 K\o benhavn \O, Denmark.}
\author{Kim Sneppen}
 \email{sneppen@nbi.ku.dk}
\affiliation{Niels Bohr Institute
}

\date{\today}
\begin{abstract}
    In population dynamics, mathematical models often contain too many parameters to be easily testable. A way to reliably estimate parameters for a broad range of systems would help us obtain clearer predictions from theory. In this paper, we examine how the allometric scaling of a number of biological quantities with animal mass may be useful to parameterise population dynamical models. Using this allometric scaling, we make predictions about the ratio of prey to predators in real ecosystems, and we attempt to estimate the length of animal population cycles as a function of mass. Our analytical and numerical results turn out to compare reasonably to data from a number of ecosystems. This paves the way for a wider usage of allometric scaling to simplify mathematical models in population dynamics and make testable predictions.
\end{abstract}


\maketitle

\section{Introduction}

When modeling the dynamics of ecological communities, a recurring problem is the difficulty of estimating model parameters. If we desire to develop a model that can describe real ecosystems, a common approach is to add terms and parameters to account for as many real-world complications as possible. The result is unfortunately that many of the models end up being too complicated to actually make any definitive predictions due to uncertainties about the often large number of parameters. A model that requires precise measurements of parameters for every individual system one wishes to study will of course be interesting for the isolated case, but it will be difficult to derive more general principles from it. We believe that a simplified model that makes approximate but clear predictions might be a more useful approach. In this paper, we will argue that, by using allometric mass scaling, it is possible to estimate the parameters of the classic Lotka-Volterra predator-prey equations in such a way that this highly idealised model can be used to predict the behaviour of actual populations. It is our hope that we will be able to rewrite all parameters of the equations in terms of only two quantities: Prey mass and predator mass. We will also look at the implications of body size for the period of animal population cycles. By doing so, we wish to conclusively demonstrate the usefulness of allometric mass scaling relations in population modeling.

The fact that many ecological variables scale allometrically with animal body mass has attracted increasing attention in recent years. Ginzburg \& Colyvan \cite{GinzburgColyvan04} go as far as to call the allometries fundamental laws of ecology, comparing them to Kepler's laws in physics. Peters \cite{Peters83} compiled a list of variables exhibiting allometric scaling, which we will make use of in this paper. For example, generation time and metabolic rate correlate with mass to powers of (approximately) 1/4 and 3/4, respectively. It is these relationships that we will exploit to write the Lotka-Volterra equations in terms of animal body mass. For a compelling attempt at finding a theoretical foundation for these quarter-power scaling laws, see the work of West et al. \cite{Westetal97}.

On the larger ecosystem scale, there are also examples of allometric scaling. In particular, many animals - most prominently rodents such as lemmings - exhibit a regular population cycle. The time elapsed between peaks in abundance of such animals tends to scale with the average mass of the animal. Empirically, the scaling relation is found to be $T\propto m^{0.26}$ \cite{Petersonetal84}. We wish to argue for a theoretical basis of this relationship.

Yodzis \& Innes \cite{YodzisInnes92} use the mass to parameterise a system of equations similar to generalised Lotka-Volterra equations, with consumer and resource (whether plant or animal) substituted for predator and prey. Their model assumes that the predator reproduction will saturate with increasing prey population, giving the predator a Holling type II or III functional response. Also, they argue that the strength of the predator-prey interaction should scale with the ratio of prey mass to predator mass to some power, and that it should be possible to determine the coefficients of this scaling law from measurable biological quantities. With the model in place, they analyse the linear stability of the dynamical system and find that, for certain predator-prey mass ratios, it will have a limit cycle with a period $T\propto m_C^{1/8} m_R^{1/8}$, where $m_C$ is the consumer (predator) mass and $m_R$ is the resource (prey) mass.

We will here proceed down a similar path, though our model will be notably simplified and our approach to the predator-prey functional response will be different. The original Lotka-Volterra equations on which we will be basing our model assume that the predation and predator reproduction rates increase in proportion with prey population density, a so-called Holling type I functional response. We here assume that prey population is always far from the carrying capacity of the ecosystem, resulting in a prey reproduction rate that is also proportional to prey population. Instead of trying to determine a biologically reasonable coefficient for the scaling of interaction strength with the predator-prey mass ratio, we will let the coefficient remain unknown. We will determine the equilibrium populations in terms of this unknown coefficient. Luckily, it turns out that when we look at the ratio of the populations, this coefficient cancels out. Thus, our method still yields useful information.

Finally, we will look at the period of population cycles. The simplest version of the Lotka-Volterra equations has a non-trivial equilibrium which is a center, rather than a limit cycle. Here, we likewise find a period of $T\propto m_C^{1/8} m_R^{1/8}$ as mentioned above. In order to obtain the empirically determined  $m^{1/4}$ relationship with population cycle length, Yodzis \& Innes point out that one can assume a direct proportionality between predator (consumer) size and prey (resource) size. While this relationship may hold in many systems (see e.g. \cite{Brose10}), it certainly does not in such cases as the wolf-moose system studied by Peterson et al. \cite{Petersonetal84}, and the relationship is hardly well-defined in systems where the resource is a plant. Ginzburg \& Colyvan \cite{GinzburgColyvan04} even present a critique of the whole idea of using only the linearisation of the Lotka-Volterra equations to predict the length of population cycles. It would therefore be preferable if we could derive a relationship between prey mass and cycle period that is independent of predator mass. This is what we will attempt to do in the following.

The model put forward here is thus an application of the basic idea of Yodzis \& Innes to a heavily simplified system of equations, without making attempts at determining the exact interaction strength between predator and prey directly. It is our hypothesis that even such a simplified model will still give reasonable order-of-magnitude predictions about real ecosystems.

\section{Parameterising the Lotka-Volterra equations}

The original Lotka-Volterra predator-prey equations read as follows:

\begin{equation}
    \frac{dx}{dt}=\alpha x - \beta x y
\end{equation}

\begin{equation}
    \frac{dy}{dt}=\gamma x y - \delta y
\end{equation}

\cite{Lotka20}. Here, $x$ denotes prey, $y$ predator, $\alpha$ is the per-capita reproduction rate of prey, and $\delta$ is the per capita death rate of predators in the absence of prey. The interaction strengths $\beta$ and $\gamma$ are slightly harder to define. $\beta$ denotes the risk of each prey being eaten per predator, and $\gamma$ represents the increase in predator reproduction rate per prey. These latter two parameters are of course more difficult to estimate than the first two, and we will therefore need to find a way around this obstacle.

As opposed to Yodzis \& Innes, we choose to work with animal abundances rather than biomass densities. We do this because it is conceptually easier and data are more readily available for abundances than for biomass densities for the systems that we wish to study. A complication arising from this is that when working with abundances, there is a distinction between somatic growth (individuals growing larger) and reproductive growth, which would be unimportant if we were to work with biomass densities. We shall therefore ignore the finer details of animal reproduction and growth, and simply assume that all growth results in the production of new individuals. Furthermore, we assume that the populations are large enough and reproductive events evenly distributed enough in time that population growth can be modelled as continuous rather than discrete.

According to Peters \cite{Peters83} we then have the following empirical relation for reproduction rate:

\begin{equation}
    \alpha = \frac{1}{400} m_x^{-1/4} \hspace{3mm} [\mathrm{day}^{-1}]
\end{equation}

 In the cited mass scaling relations, all masses are in kilograms. As the predator-prey pairs we will be examining here are all mammals, we shall be using the mass scaling relations that apply to mammals. For cold-blooded animals such as reptiles the relations will be different, though not radically so.

It should be possible to calculate the death rate of predators in the absence of prey from the so-called turnover time. This is defined as the average time it will take an animal to metabolise its entire energy reserves. In turn, this can be calculated from the metabolic effect. Again from \cite{Peters83}

\begin{equation}
    t_{turnover} = 19 m_y^{1/4} \hspace{3mm} \mathrm{[day]}
\end{equation}

This implies

\begin{equation}
    \delta = t_{turnover}^{-1} = \frac{1}{19} m_y^{-1/4} \hspace{3mm} \mathrm{[day^{-1}]}
\end{equation}

The coupling coefficient $\beta$ we assume to be proportional to predator ingestion rate. We believe this to be justified, since the more a given predator consumes, the higher the per-capita risk of being eaten by it should be for the prey. The predator ingestion rate in terms of energy scales with mass as

\begin{equation}
    I \propto m_y^{3/4} \hspace{3mm} [\mathrm{J \cdot (day \cdot predator)^{-1}}]
\end{equation}

\cite{Peters83}. The number of individual prey that a predator needs to eat to satisfy this energetic demand is inversely proportional to prey mass, and we therefore write $\beta$ as

\begin{equation} \label{eq:beta}
    \beta = k \cdot \frac{ m_y^{3/4}}{m_x} \hspace{3mm} [\mathrm{(day \cdot predator)^{-1}}]
\end{equation}

where $k$ is an unknown proportionality constant. Knowing the equilibrium population of prey or predator should make it possible to determine $k$ if this is desired. 

Our parameterisation thus deviates notably from that of Yodzis \& Innes, since they assume that the predator death rate and the interaction strength scale with the ratio of prey mass to predator mass to the power of 3/4 (here converted to abundance rather than biomass, as was originally used). Strictly speaking, the ingestion rate of $y$ predators reflects some kind of average prey consumption rate at average prey abundance. What we really need here is the slope of predator kill rate as a function of prey abundance. Furthermore, the units of the ingestion rate is $\mathrm{[J \cdot (predator \cdot day)^{-1}] \propto [prey \cdot (predator \cdot day)^{-1}]}$ and not $\mathrm{[(predator \cdot day)^{-1}]}$ as we need it to be for our units to match. Despite all this, we still believe that the allometric scaling of the ingestion rate is a reasonable approximate measure of the predator's ability to consume and therefore of the dependence of consumption rate on prey abundance. We now only need to find a way around not knowing the exact proportionality.

The slope of predator kill rate with prey abundance that really constitutes $\beta$ depends on a number of factors (temperature, prey population density, predator satiation etc. \cite{Peters83}), and it is probably not possible to make a universal estimate of it. Instead, we let $k$ embody all these complications and tune it to fit the systems that we will study. As mentioned above, it fortunately cancels out in the final calculation of the prey to predator population ratio anyway.

The relation between the number of prey eaten and the number of predators produced can be derived approximately if we know the ecological efficiency $\eta$ of the predator-prey interaction. The ecological efficiency here refers to the percentage of prey biomass that is converted into predator biomass. Ecological efficiencies vary considerably depending on the nature of the interaction \cite{Lindeman42}, and it is therefore difficult to find an estimate that is both precise and general. For systems with a low predator to prey mass ratio and positive correlation of biomass density with body mass, ecological efficiency should be high ($\eta \approx 35$ \%) according to a review by Trebilco et al.\cite{Trebilcoetal13}, which, however, deals with aquatic ecosystems. Lindeman's original paper similarly shows an efficiency that rises with trophic level \cite{Lindeman42}. On the other hand, a case study of the Isle Royale wolf-moose system that we will discuss below suggests that the wolves have a much lower efficiency than we would expect based on the above ($\eta \approx 2$ \%) \cite{Colinvaux79}. In laboratory experiments, a figure of about $\eta = 10$ \% is observed \cite{Colinvaux79}, and for lack of a better estimate, we shall use this so-called ten percent law in our calculations. Given that we are not going for an exact description of any one particular interaction, we believe that it is justified to use this rough estimate.

The relation between mass of consumed prey ($m_{x,c}$) and mass of produced predator ($m_{y,p}$) is now

\begin{equation}
    m_{y,p} = \eta \cdot m_{x,c}
\end{equation}

assuming that prey and predator have similar energy content per unit mass. Rewriting this in terms of numbers of individual predators produced ($N_{y,p}$) and prey consumed ($N_{x,c}$) we get

\begin{equation}
    N_{y,p} = m_{y,p}/m_y = \frac{m_{x,c}}{m_y}\eta =  \frac{m_x N_{x,c}}{m_y}\eta
\end{equation}

In the Lotka-Volterra equations, the number of predators produced per unit time is given by the term

\begin{equation}
    N_{y,p} = \gamma x y
\end{equation}

and the number of prey consumed  by the term

\begin{equation}
    N_{x,c} = \beta x y
\end{equation}

Thus, we get the following relation between $\beta$ and $\gamma$ :

\begin{equation}
    \gamma = \frac{m_x}{m_y} \eta \cdot \beta =  k \cdot m_y^{-1/4} \cdot \eta \hspace{3mm} \mathrm{[(day \cdot prey)^{-1}]}
\end{equation}

We have now written all the parameters of the equations in terms of the animal body masses alone, with $k$ from eq. (\ref{eq:beta}) being the only parameter that remains to be determined. However, we can get around this by focusing our attention on the equilibrium predator-to-prey population ratio instead of the absolute populations.

The Lotka-Volterra equations have the non-trivial equilibrium 

\begin{equation}
    (x,y) = \left(\frac{\delta}{\gamma} \hspace{1mm}, \hspace{1mm} \frac{\alpha}{\beta} \right)
\end{equation}

 which is neutrally stable. The equilibrium ratio between prey and predator populations is therefore

\begin{equation} \label{eq:Ratio}
    x/y = \frac{\beta \delta}{\alpha \gamma} = \frac{21}{\eta} \left(\frac{m_y}{m_x}\right)^{3/4}
\end{equation}

This number depends only on the masses. We see that due to the factor $1/\gamma$ this ratio is inversely proportional to ecological efficiency, so that if our estimated 10 \% efficiency is a factor 2 too great, we will estimate a ratio that is half the "correct" value.

\section{The period of population cycles}

\begin{figure}
\includegraphics[width=0.39\paperwidth]{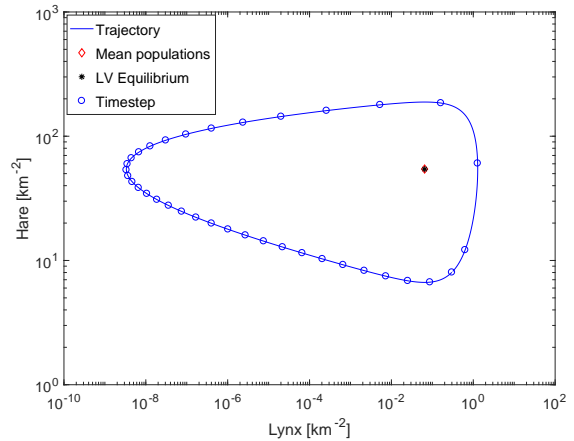}
\caption{(Colour online) An illustration of the dynamics of the predator-prey system. This numerical solution is based on parameters appropriate for the lynx-hare system discussed below. The line shows the trajectory of the system in predator-prey space, and the circles are all spaced evenly in time, at a separation of 50 days. The distinction between a fast and a slow segment of the trajectory can be clearly seen from the spacing of the circles. Note also that equilibrium abundances are practically identical to mean abundances, meaning that we can use the two interchangeably.}
\label{fig:Trajectory}
\end{figure}

The Jacobian matrix of the Lotka-Volterra equations at the non-trivial steady state has the eigenvalues $(i\sqrt{\alpha \delta} , -i \sqrt{\alpha \delta} )$, meaning that for small perturbations away from equilibrium, the system will oscillate over time with a period of $T=\frac{2 \pi}{\sqrt{\alpha \delta}}$. This leads to the aforementioned scaling of population cycle period with mass $T \propto m_x^{1/8} m_y^{1/8}$, contrary to the observed $T \propto m_x^{1/4}$. A problem with using linearisation in this case is that the period thus obtained only applies when oscillations are relatively small. Population cycles in actual predator-prey pairs, such as the vole-weasel pair in northern Scandinavia, can involve fluctuations over two orders of magnitude \cite{Hanskietal01}. When solving the equations numerically, we see that much of the time, the population of prey will be in a state of slow, exponential recovery, while the predator population slowly approaches zero. When the prey population recovers, the predator population quickly explodes, initiating a swift collapse of the prey population. The collapse phase observed in real rodent cycles does indeed appear to be notably shorter than the growth and peak phases, and the corresponding predator cycles are similarly observed to be very sharply peaked \cite{Krebsetal95, Hanskietal01}. We therefore believe that the dynamics can be realistically modelled as consisting of a slow exponential growth phase and a fast collapse phase. Using this two-timescale assumption, we will try to derive an expression for the period $T$ of population cycles. Splitting more complex predator-prey models into slow and fast phases has previously been done by Rinaldi \& Muratori \cite{RinaldiMuratori92}. In the following, we shall use a similar basic idea, but a different mathematical approach and solve for the period $T$, rather than maximal abundance as they did. An illustration of the of the cycle and its fast and slow segments can be seen in fig. (\ref{fig:Trajectory}). For our derivation, we will use the maximum and minimum prey density of a cycle, which should be easily obtainable from observations and available in the literature.

The slow approach to and subsequent drifting away from the saddle point at $(0,0)$ is what takes up the majority of the orbital period of the system. For this reason, we will here attempt to derive an approximate relation for the cycle length by looking at the behaviour around the saddle point at $(0,0)$ instead of the center at $\left( \frac{\delta}{\gamma} \hspace{1mm}, \hspace{1mm} \frac{\alpha}{\beta} \right)$. Although the period of the cycle is mainly determined by the hyperbolic approach to the saddle point, the oscillation still happens around the center equilibrium at $\left( \frac{\delta}{\gamma} \hspace{1mm}, \hspace{1mm} \frac{\alpha}{\beta} \right)$. As can be seen in fig. \ref{fig:Trajectory}, the time average populations are very close to the equilibrium populations at the center. We therefore do not believe that there is a contradiction between using the saddle point linearisation to determine the oscillation period, but determining population ratios based on center equilibrium values. 

Using the linearisation around the center equilibrium, we obtain a period that is independent of initial conditions, but which does not match observations, as the assumption that initial conditions are close to the equilibrium breaks down in the real systems studied here. Instead, we assume that the initial conditions are far from the center equilibrium. For this asymptotic approximation, the period will depend on initial conditions and the calculated period matches observations better.

\begin{figure*}
(a)\includegraphics[width=0.39\paperwidth]{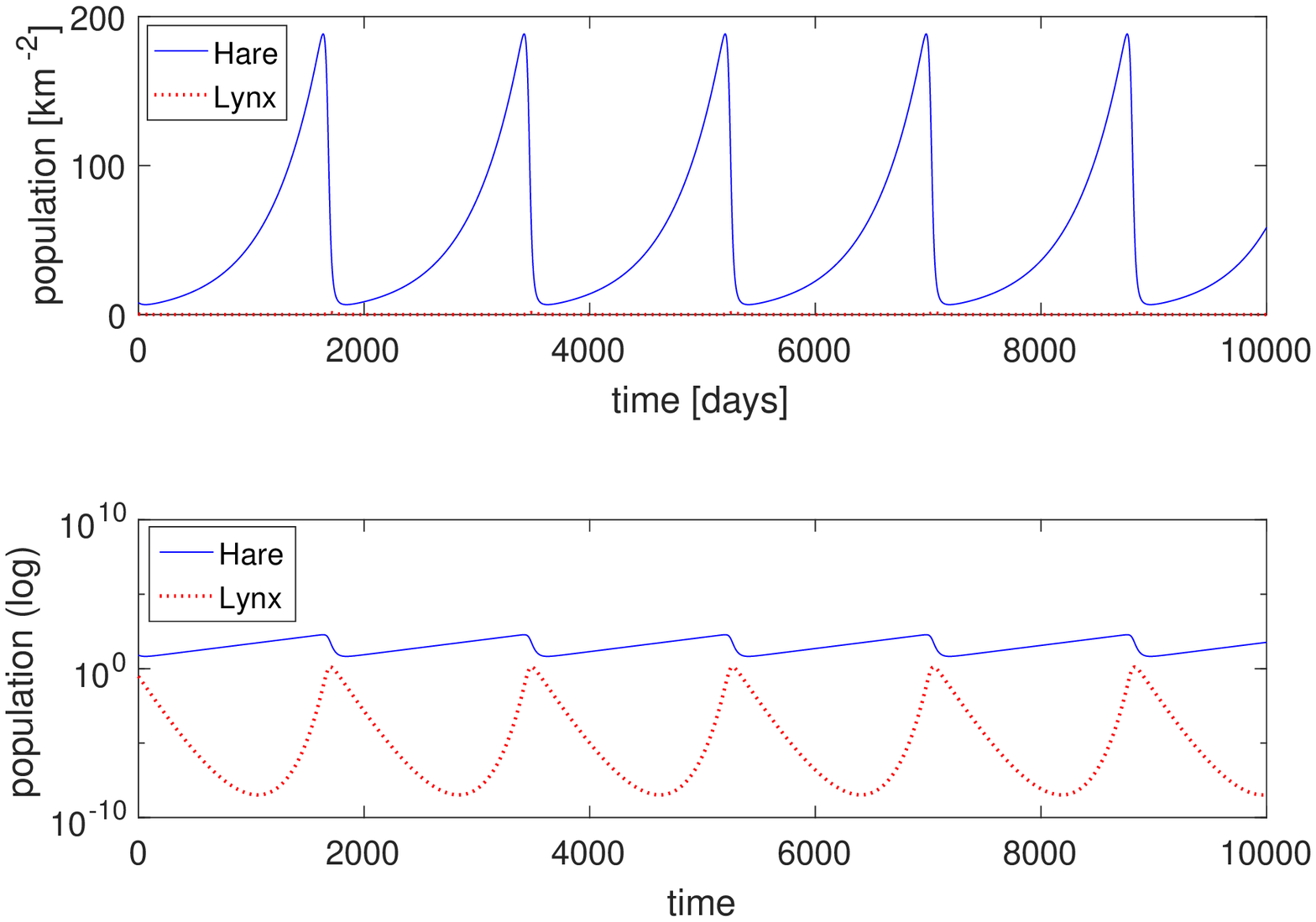}
(b)\includegraphics[width=0.39\paperwidth]{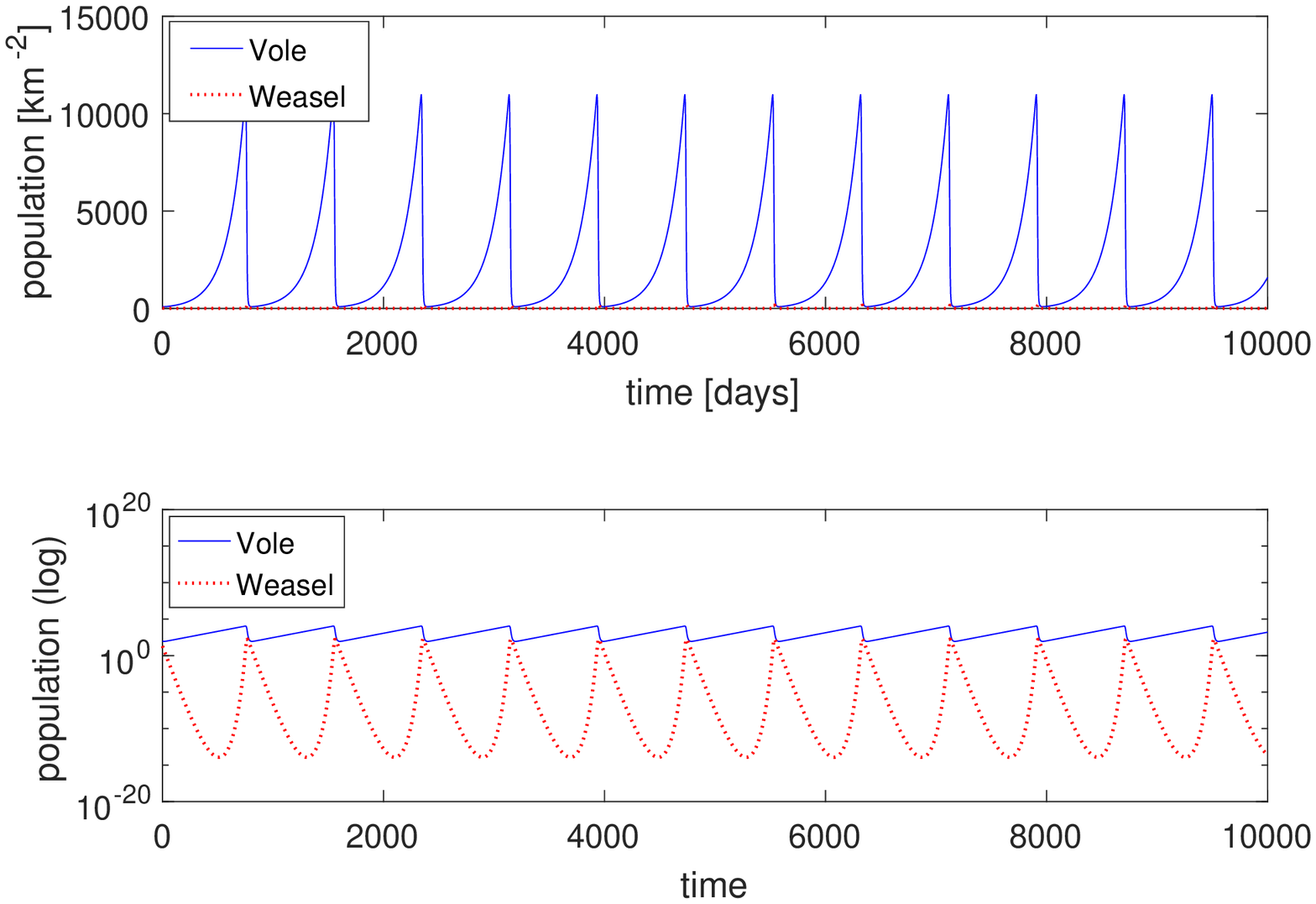}
\caption{(Colour online) (a) A numerical simulation of the Lotka-Volterra equations for lynx and hare. $x_{max} \approx 180$ km$^{-2}, x_0 = x_{min} \approx 8 $km$^{-2}$, $y_0 = 0.3$ and $k=1.05 \cdot 10^{-2}$. The period is just under 2000 days, or 5.5 years, and the average hare density is 51 km$^{-2}$. Average lynx density is 0.059 km$^{-2}$. The ratio of the averages is 860 hares per lynx. As can be seen from the logarithmic plot, the predicted predator oscillations are too violent, with extinction of lynx at the cycle minimum. When this does not actually happen, it may be due to the fact that lynx can survive partially on other prey when hare population is low \cite{SunquistSunquist02}. (b) The solution obtained using the masses of voles and weasels. $x_{max}\approx 10^4$ km$^{-2}$, $x_0 = x_{min} \approx 10^2$ km$^{-2}$, $y_0 = 20$, and $k=2 \cdot 10^{-4}$. We still see a cycle somewhat shorter than the observed, with an estimated $T \approx 2.3 \hspace{1mm} \mathrm{years}$. Again, the predator oscillation is unrealistically violent. Average vole density is 2100 km$^{-2}$ and weasel density is 4.6 km$^{-2}$, giving 460 voles/weasel.}
\label{fig:Numerical}
\end{figure*}

Starting from a population $x_{min}$, the prey population should grow as follows:

\begin{equation}
    x(t) = x_{min} \mathrm{e}^{\alpha t}
\end{equation}

When predator population is low, prey population grows unobstructed. After one period of length $T$, we should have the maximal population density

\begin{equation}
    x_{max} = x(T) = x_{min} \mathrm{e}^{\alpha T}
\end{equation}

The time it will take the population to recover to a density of $x_{max}$ now becomes

\begin{equation} \label{eq:T}
    T = \frac{1}{\alpha}\ln\left(\frac{x_{max}}{x_{min}}\right) = 400\cdot \ln\left(\frac{x_{max}}{x_{min}}\right) \cdot m_x^{1/4} \hspace{3mm} \mathrm{[day]}
\end{equation}

We thus get the $m_x^{1/4}$-relation found empirically. The above expression should be valid when the amplitude of oscillations is very large, so that the period of the predator-prey cycle is dominated by the slow growth phase which in the Lotka-Volterra model occurs at low predator abundances. Note, however, that we at no point have assumed that the population crash should be due to the influence of a predator. We just assumed that the crash was fast and did not extend the period length or influence the exponential growth phase significantly. The derivation here should therefore be equally valid if a population crash is caused by e.g. a shortage of food or an epidemic. Given that in the case of many rodents it is unclear if it is actually predation that drives the cycle \cite{Turchin00}, this is a significant advantage.

Another interesting feature of this expression is the logarithmic scaling with population maximum-minimum ratio. Hanski has already hinted at such a scaling relation for the vole-weasel system \cite{Hanski91}. In his 1991 paper, he shows that $\ln\left(\frac{x_{max}}{x_{min}}\right)$ correlates with latitude, and that oscillation period also correlates with latitude. Oscillation period thus also correlates with the logarithm of the maximum-minimum ratio. It is possible that we have found a theoretical explanation for this correlation.

In the next section, we will demonstrate that eq. (\ref{eq:T}) roughly fits the pattern seen in oscillating populations in nature, although there is a significant deviation between predicted and observed numbers. For the prey-predator ratios on the other hand, the parameters derived above mostly give realistic results.

\begin{table*} 
\begin{center}
\begin{tabular}{|l|c|c|c|c|c|c|c|c|}
    \hline
     System  & $m_y$ [kg] & $m_x$ & $x_{max}$ & $x_{min}$ & Observed $x/y$ ratio & Theoretical $x/y$ & Obs. $T$ [days] & Theoretical $T$ \\ \hline
     Lynx-hare & $11 \pm 1$ & $1.6 \pm 0.1$ & $180 \pm 80$ & $8 \pm 4$ & $600 \pm 400$ & $850 \pm 70$ & $3000 \pm 200$ & $1400 \pm 300$ \\ \hline
     Vole-weasel & $0.08 \pm 0.01$ & $0.025 \pm 0.002$ & $(10 \pm 2) \cdot 10^4$ & $100 \pm 50$ & $200 \pm 200$ & $510 \pm 60$ & $1600 \pm 200$ & $730 \pm 90$ \\ \hline
     Wolf-moose & $33 \pm 1$ & $350 \pm 10$ & - & - & $40 \pm 20$ & $42 \pm 2$ & - & -  \\ \hline
     Lemming osc. & - & $0.064 \pm 0.003$ & $1000 \pm 200$ & $14 \pm 5$ & - & - & $1460 \pm 0$ & $860 \pm 80$  \\
    \hline
\end{tabular}
\end{center}
\caption{Table of the data used and the values calculated, including uncertainties. Numbers are rounded to the highest uncertain digit \cite{Hanski91,Gilg03,Henttonen87,Carlsen99,Moen10,Gillingham84,Seal83}. \label{Tab:Data}}
\end{table*}

\section{Comparing theory with data}
The classic example of a system described well by the Lotka-Volterra equations is the interaction between the canadian lynx (\emph{lynx canadensis}) and the snowshoe hare (\emph{lepus americanus}). Although there has been some doubt as to whether the hare population cycle is driven primarily by predation or other factors, there seems to be evidence that changes in hare mortality are mainly due to predation \cite{Krebsetal01}. The population density of hares oscillates from around 8 to just under 200 per square kilometer over the 8-10 years long cycle \cite{Krebs11}. The average density of lynx ranges from 0.03 to 0.3 km$^{-2}$ \cite{Mowatetal00}.

To see how well our model fits with observations, we plug the average masses of lynx - on average roughly 11 kg \cite{Moen10} and hares - roughly 1.6 kg \cite{Smithetal80} into the equations and solve them numerically. We choose initial conditions corresponding to the density per square kilometer when hare abundance is lowest ($x_0 \approx x_{min} = 8$ and $y_0 = 0.3$ - due to the phase difference between lynx and hare population oscillations, we let lynx population start out high and hare population start out low). We then tune the parameter $k$ to obtain the correct ratio between cycle highs and lows. Initial lynx abundance is taken to be slightly above minimum, as the predator cycle will lag behind the prey. The result can be seen in fig. \ref{fig:Numerical} (a). Our simulation predicts an average prey to average predator population ratio that is quite close to the observed values. The period is off by about a third, which, given the simplifications of the model is not a bad estimate. The fact that the population collapse takes such a short time in our simulation contributes to our underestimating the period. In reality, the collapse takes about 1-2 years \cite{Krebs11}. The spiky appearance of the graph is also not very naturalistic. However, taking increasing predation from other predators, increasing susceptibility to disease and other complicating factors that increase with population density into account would most likely lead to a more rounded shape of the peaks, similar to the one seen in actual observations. It turns out that the time average abundances are fairly close to the predicted equilibrium abundances in all of our numerical solutions. We shall therefore use mean abundances and equilibrium abundances interchangeably when validating our results.

We also plug the masses into eqs. (\ref{eq:Ratio}) and (\ref{eq:T}). The theoretical estimates obtained this way and their uncertainties can be seen in table \ref{Tab:Data}. For this particular system, we estimate a period of $T \approx 1400 \hspace{1mm} \mathrm{days}$. Compared to the observed period of around 3000 days, the error is about 50 \%. As far as order-of-magnitude estimates go, this is still reasonable. Neglecting the duration of the collapse phase is probably part of the reason for this error. For comparison, the cycle period obtained from linearisation gives us $\frac{2\pi}{\sqrt{\alpha \delta}} = \frac{550}{(m_x m_y)^{1/8}} = 770 \hspace{1mm} \mathrm{days}$, which is far too short. This again underlines the usefulness of approximating the cycle as a series of instantaneous collapse phases followed by exponential growth phases.

Another case where the basic Lotka-Volterra equation might be useful is the interaction between the vole (\emph{microtus agrestis}) and least weasel (\emph{mustela nivalis}) in northern Scandinavia, as mentioned above. Although there still is some doubt about the role of predation in the cycle here as well, there is evidence that predation plays at least a significant part. Vole density ranges from $10^2$ to $10^4$ \hspace{1mm} km$^{-2}$ over a cycle, while weasel density ranges from 1 to 20 km$^{-2}$ and is strongly correlated with vole density at northern latitudes \cite{Hanski91}. The cycle is observed to be about 4 years long in the areas we are interested in \cite{Hanskietal01}. A numerical solution of the Lotka-Volterra equations for these parameter values can be seen in fig. \ref{fig:Numerical} (b). This numerical solution gives us an estimate of the period $T \approx 830 \hspace{1mm} \mathrm{days} = 2.3 \hspace{1mm} \mathrm{years}$, and of the prey-predator ratio of 460 voles per weasel. A comparison between theoretical results calculated using the derived expressions and observations can again be seen in table \ref{Tab:Data}.

Large population oscillations are observed in some rodent species even when there is no single obvious predator feeding on the rodent. One example of this is the northern collared lemming of Greenland (\emph{dicrostonyx groenlandicus}) \cite{Gilg03}. We of course cannot use such an example to test our hypothesis about prey-predator population ratios, but we may still use it to examine the accuracy of the derived period. The results of our examination can be seen in the table, and both the estimated period and the error are similar to those of the vole.

\begin{figure*}
    \centering
    (a)\includegraphics[width=0.39\paperwidth]{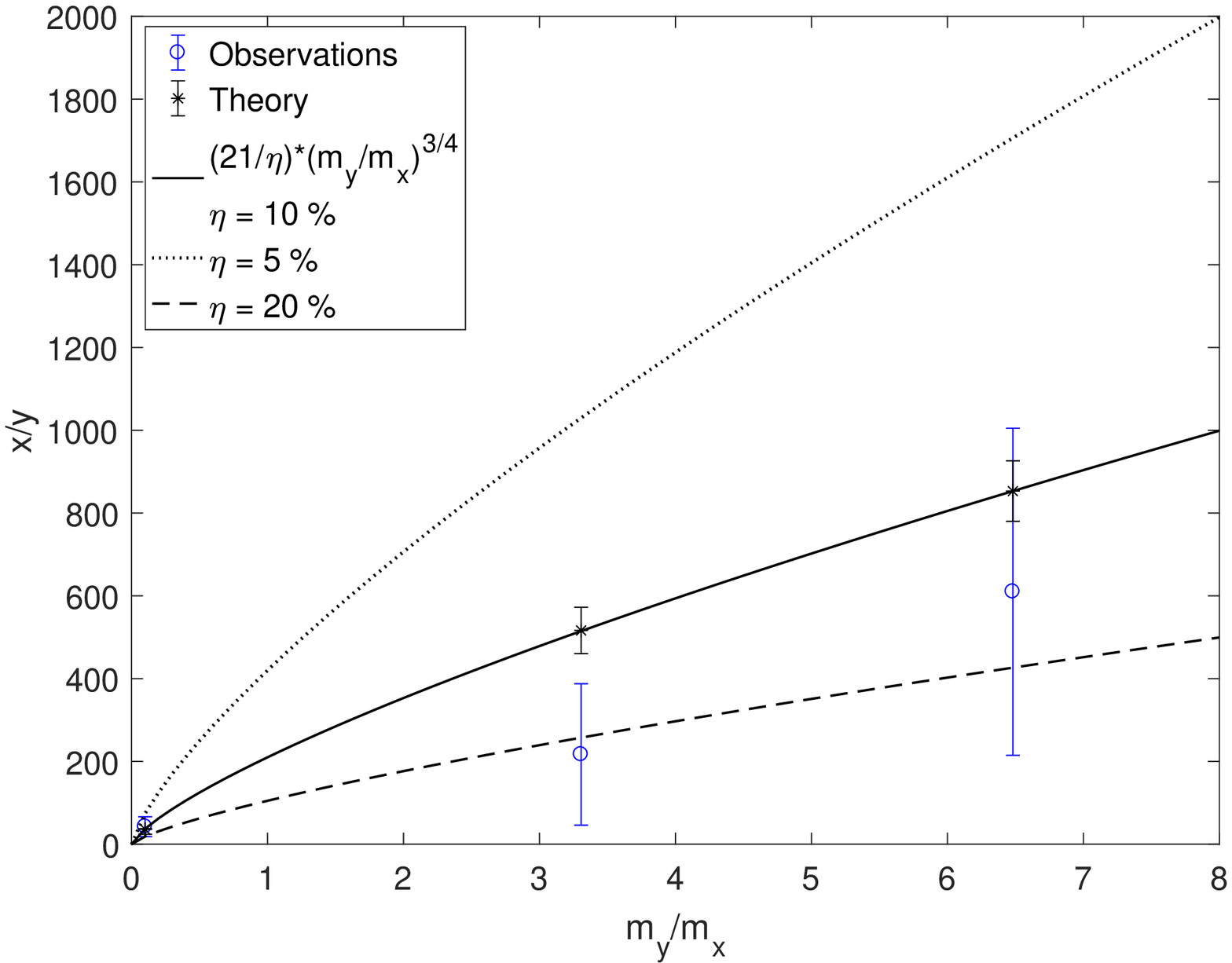}
    (b)\includegraphics[width=0.39\paperwidth]{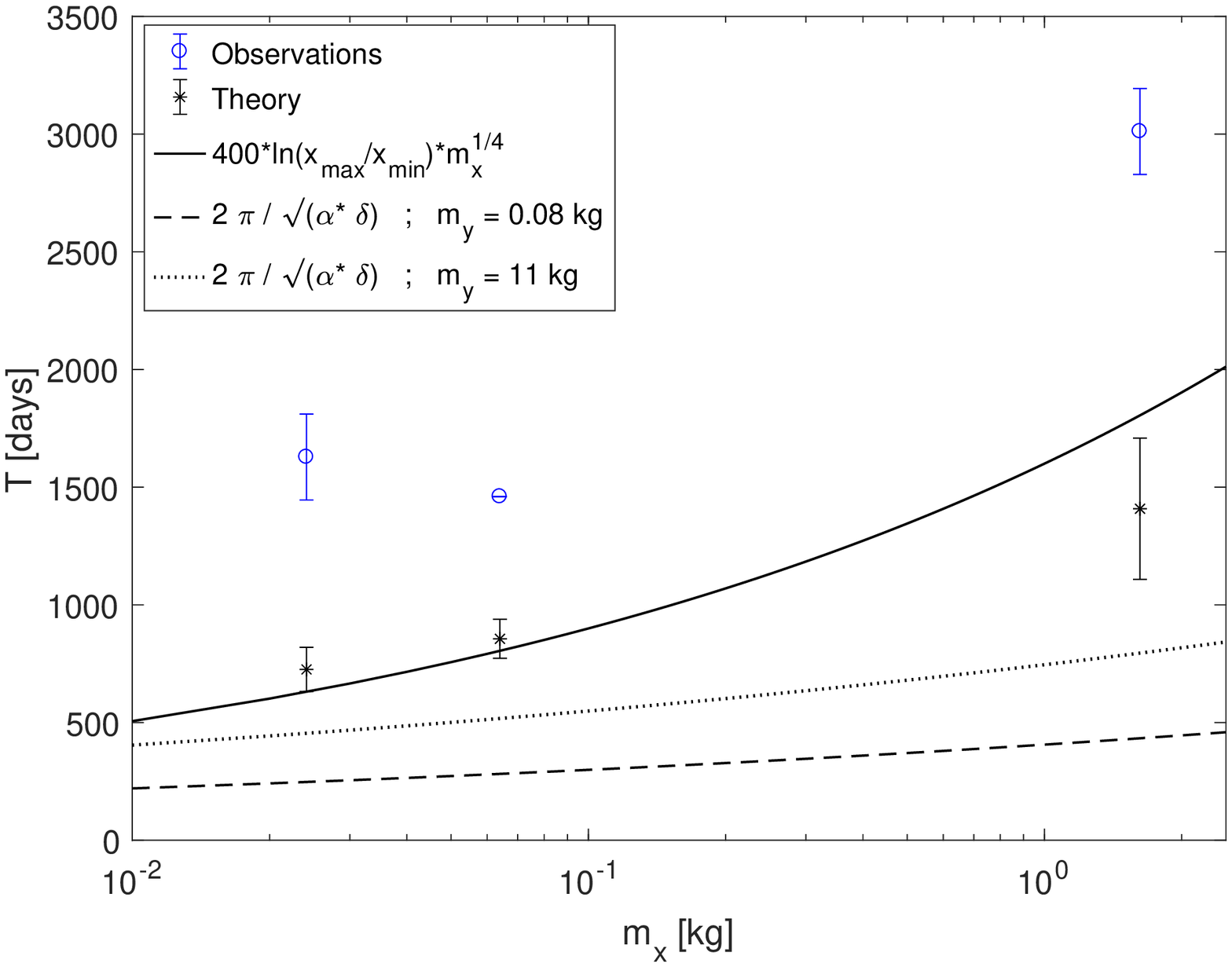}
    \caption{(Colour online) (a) shows the observed and theoretically calculated prey-predator population ratios for the wolf-moose, vole-weasel, and hare-lynx systems. Here, the full line shows the predicted power law. The dotted line shows the theoretically calculated prey-predator ratio for half the ecological efficiency used in this paper (5 \%), while the dashed line shows the prey-predator ratio calculated for twice the used ecological efficiency (20 \%). (b) shows the observed and calculated periods of population oscillations for voles, lemmings, and hares. The black line shows the corresponding mass power law where we have set $\ln \left(\frac{x_{max}}{x_{min}}\right) = 4$. This number is close to the values for the vole and lemming oscillation. The dotted and dashed lines show the $(m_x m_y)^{1/8}$ scaling law predicted from linearisation, where $m_y$ is that of lynx and weasel respectively.  We could not calculate an oscillation period for the moose of Isle Royale, and no single predator is known to cause the lemming oscillation, so they each only occur in one of the plots. Data points and error bars show the numbers without any rounding.}
    \label{fig:Errorbars}
\end{figure*}

As a final example, we will consider the wolves (\emph{canis lupus}) and moose (\emph{alces alces americanus}) of Isle Royale in Lake Superior, Michigan. On this island, wolves and moose coexist in isolation, with very little interference from other animals. Due to the small size of the island, animal populations are so small that random events (such as the introduction of parvovirus to the wolf population in 1980) will have a large influence on the population, which seems to fluctuate almost erratically \cite{VucetichPeterson04}. Therefore, we cannot determine an observational population cycle length for this system. However, the average populations should reflect an equilibrium ratio that should be predictable from wolf and moose mass. As can be seen in table \ref{Tab:Data}, we obtain an accurate estimate of this ratio.

Based on these cases we may conclude that our idealised model works as an order-of-magnitude estimate of the behaviour of ecosystems. There is a discrepancy between the derived period of population oscillations and what is observed, and this discrepancy cannot be explained entirely by experimental uncertainty. However, our results reproduce two patterns observed empirically, which have not yet been theoretically explained. One is the apparent scaling of oscillation periods with mass to the quarter power. Another is the scaling of period with $\ln\left(\frac{x_{max}}{x_{min}}\right)$. We will therefore argue that the derived expression is of interest despite the discrepancy.

\section{Discussion}

As predicted in the introduction, we have been able to parameterise the Lotka-Volterra equations using animal body mass in such a way that they provide fairly accurate predictions of the equilibrium predator-prey population ratio. When we also know the amplitude of the fluctuations of prey population, we obtain analytical estimates of the oscillation periods that reproduce the patterns found in nature, albeit with a discrepancy. Notably, our approximate expression for the cycle period exhibits the same allometric mass scaling as the one found empirically. Furthermore, it shows a logarithmic scaling of period with the ratio of maximum to minimum populations, which is also found in data. The ratios of average prey population to average predator population found in our simulations fit relatively well with real-world data. For the population ratios, the uncertainty of population counts and animal weights explain the errors in two of three cases. Our prediction of the amplitude of predator oscillations, however, is unreasonable in comparison with observations, possibly because of the assumption that the predator is entirely dependent on one prey species.

Of course, even though our model was only meant as a crude estimate, we need to address why we see the discrepancy that we do between theory and observations. In the case of the prey-predator population ratios, the uncertain estimate of ecological efficiency is a likely source of error. The range of ecological efficiencies observed in the real world is so large that it poses a challenge to this kind of population dynamical modelling. If our estimated efficiency is a bit too low, it will explain the discrepancy, as can be seen in fig. \ref{fig:Errorbars}.

The period is off by a larger percentage, and it is less clear what might cause the error. One drastic assumption that we have made is that it takes no time for animals to grow to adult size. We have considered whether this delay might explain some of the error. To take the time required to reach full size into account, we have attempted a numerical solution of the equations while including a time delay in predator and prey reproduction. Unfortunately, this does not significantly change the oscillation period. Another possible source of error is the assumption that collapse is instantaneous. In reality, it does take some time, though not as long as the exponential growth phase. If the duration of the collapse phase also scales with animal mass, it would help explain why we consistently underestimate the period by about 50 \%. 

Finally, our model is a mean-field theory, whereas in reality, geographical separation does play a role. Maybe the fact that real predators have to seek out the prey, and that prey may survive for longer in some locations than in others may serve to slow the dynamics of real, geographically extended ecosystems. This, however, is a subject that we will leave for future studies.

Despite these discrepancies, our work demonstrates that, by using the many available allometric mass scaling laws, it is possible to obtain reasonable predictions from even very simple population dynamical models. This fact should have wide applications in population dynamics. Another area where this could be applicable is in epidemiology. The incubation and recovery times of a variety of diseases with multiple host species have already been shown to scale with host mass \cite{Cableetal07}, and Dobson \cite{Dobson04} has studied a multi-host disease model parameterised using mass. A possible further use of the model described here could be to construct an epidemiological model taking predation into account. Models of epidemics in predator-prey systems have been proposed before \cite{HsiehHsiao08}, but they often contain so many unknown parameters that an examination of parameter space becomes difficult. Here, a parameterisation using mass could significantly reduce the number of free parameters.

Already in 1992, Yodzis \& Innes pointed out  that the application of mass scaling relations to population dynamics can potentially make it a lot easier to make realistic estimates of the parameters involved. Still, to our knowledge it is not until now that the predictions of a mass-parameterised population dynamical model have been tested against real-world data. The scaling of reproduction rate with animal mass has also provided us with a possible explanation for the relationship between population cycle length and mass, at least in systems where the amplitude is large. This is for example very much the case for several rodent and lagomorph species. In conclusion, we find that the allometric mass scaling laws that apply to a variety of biological quantities could potentially prove highly useful in population dynamics.

\section*{Acknowledgments}
The authors wish to thank Dr. David Vasseur for his detailed and useful comments on an earlier draft of this article.


\begin{thebibliography}{30}%
\makeatletter
\providecommand \@ifxundefined [1]{%
 \@ifx{#1\undefined}
}%
\providecommand \@ifnum [1]{%
 \ifnum #1\expandafter \@firstoftwo
 \else \expandafter \@secondoftwo
 \fi
}%
\providecommand \@ifx [1]{%
 \ifx #1\expandafter \@firstoftwo
 \else \expandafter \@secondoftwo
 \fi
}%
\providecommand \natexlab [1]{#1}%
\providecommand \enquote  [1]{``#1''}%
\providecommand \bibnamefont  [1]{#1}%
\providecommand \bibfnamefont [1]{#1}%
\providecommand \citenamefont [1]{#1}%
\providecommand \href@noop [0]{\@secondoftwo}%
\providecommand \href [0]{\begingroup \@sanitize@url \@href}%
\providecommand \@href[1]{\@@startlink{#1}\@@href}%
\providecommand \@@href[1]{\endgroup#1\@@endlink}%
\providecommand \@sanitize@url [0]{\catcode `\\12\catcode `\$12\catcode
  `\&12\catcode `\#12\catcode `\^12\catcode `\_12\catcode `\%12\relax}%
\providecommand \@@startlink[1]{}%
\providecommand \@@endlink[0]{}%
\providecommand \url  [0]{\begingroup\@sanitize@url \@url }%
\providecommand \@url [1]{\endgroup\@href {#1}{\urlprefix }}%
\providecommand \urlprefix  [0]{URL }%
\providecommand \Eprint [0]{\href }%
\providecommand \doibase [0]{http://dx.doi.org/}%
\providecommand \selectlanguage [0]{\@gobble}%
\providecommand \bibinfo  [0]{\@secondoftwo}%
\providecommand \bibfield  [0]{\@secondoftwo}%
\providecommand \translation [1]{[#1]}%
\providecommand \BibitemOpen [0]{}%
\providecommand \bibitemStop [0]{}%
\providecommand \bibitemNoStop [0]{.\EOS\space}%
\providecommand \EOS [0]{\spacefactor3000\relax}%
\providecommand \BibitemShut  [1]{\csname bibitem#1\endcsname}%
\let\auto@bib@innerbib\@empty
\bibitem [{\citenamefont {Ginzburg}\ and\ \citenamefont
  {Colyvan}(2004)}]{GinzburgColyvan04}%
  \BibitemOpen
  \bibfield  {author} {\bibinfo {author} {\bibfnamefont {Lev~R.}\ \bibnamefont
  {Ginzburg}}\ and\ \bibinfo {author} {\bibfnamefont {Mark}\ \bibnamefont
  {Colyvan}},\ }\href@noop {} {\emph {\bibinfo {title} {Ecological orbits: how
  planets move and populations grow}}}\ (\bibinfo  {publisher} {Oxford
  University Press},\ \bibinfo {year} {2004})\BibitemShut {NoStop}%
\bibitem [{\citenamefont {Peters}(1983)}]{Peters83}%
  \BibitemOpen
  \bibfield  {author} {\bibinfo {author} {\bibfnamefont {Robert~Henry}\
  \bibnamefont {Peters}},\ }\href@noop {} {\emph {\bibinfo {title} {The
  ecological implications of body size}}}\ (\bibinfo  {publisher} {Cambridge
  University Press},\ \bibinfo {address} {Cambridge},\ \bibinfo {year}
  {1983})\BibitemShut {NoStop}%
\bibitem [{\citenamefont {West}\ \emph {et~al.}(1997)\citenamefont {West},
  \citenamefont {Brown},\ and\ \citenamefont {Enquist}}]{Westetal97}%
  \BibitemOpen
  \bibfield  {author} {\bibinfo {author} {\bibfnamefont {Geoffrey~B.}\
  \bibnamefont {West}}, \bibinfo {author} {\bibfnamefont {James~H.}\
  \bibnamefont {Brown}}, \ and\ \bibinfo {author} {\bibfnamefont {Brian~J.}\
  \bibnamefont {Enquist}},\ }\bibfield  {title} {\enquote {\bibinfo {title} {A
  general model for the origin of allometric scaling laws in biology},}\
  }\href@noop {} {\bibfield  {journal} {\bibinfo  {journal} {Science}\ }\textbf
  {\bibinfo {volume} {276}},\ \bibinfo {pages} {122--126} (\bibinfo {year}
  {1997})}\BibitemShut {NoStop}%
\bibitem [{\citenamefont {Peterson}\ \emph {et~al.}(1984)\citenamefont
  {Peterson}, \citenamefont {Page},\ and\ \citenamefont
  {Dodge}}]{Petersonetal84}%
  \BibitemOpen
  \bibfield  {author} {\bibinfo {author} {\bibfnamefont {R.O.}\ \bibnamefont
  {Peterson}}, \bibinfo {author} {\bibfnamefont {R.E.}\ \bibnamefont {Page}}, \
  and\ \bibinfo {author} {\bibfnamefont {K.M.}\ \bibnamefont {Dodge}},\
  }\bibfield  {title} {\enquote {\bibinfo {title} {Wolves, moose, and the
  allometery of population cycles},}\ }\href@noop {} {\bibfield  {journal}
  {\bibinfo  {journal} {Science}\ }\textbf {\bibinfo {volume} {224}},\ \bibinfo
  {pages} {1350--1352} (\bibinfo {year} {1984})}\BibitemShut {NoStop}%
\bibitem [{\citenamefont {Yodzis}\ and\ \citenamefont
  {Innes}(1992)}]{YodzisInnes92}%
  \BibitemOpen
  \bibfield  {author} {\bibinfo {author} {\bibfnamefont {P.}~\bibnamefont
  {Yodzis}}\ and\ \bibinfo {author} {\bibfnamefont {S.}~\bibnamefont {Innes}},\
  }\bibfield  {title} {\enquote {\bibinfo {title} {Body size and
  consumer-resource dynamics},}\ }\href@noop {} {\bibfield  {journal} {\bibinfo
   {journal} {The American Naturalist}\ }\textbf {\bibinfo {volume} {139}},\
  \bibinfo {pages} {1151--1175} (\bibinfo {year} {1992})}\BibitemShut {NoStop}%
\bibitem [{\citenamefont {Brose}(2010)}]{Brose10}%
  \BibitemOpen
  \bibfield  {author} {\bibinfo {author} {\bibfnamefont {Ulrich}\ \bibnamefont
  {Brose}},\ }\bibfield  {title} {\enquote {\bibinfo {title} {Body-mass
  constraints on foraging behaviour determine population and food-web
  dynamics},}\ }\href@noop {} {\bibfield  {journal} {\bibinfo  {journal}
  {Functional Ecology}\ }\textbf {\bibinfo {volume} {24}},\ \bibinfo {pages}
  {28--34} (\bibinfo {year} {2010})}\BibitemShut {NoStop}%
\bibitem [{\citenamefont {Lotka}(1920)}]{Lotka20}%
  \BibitemOpen
  \bibfield  {author} {\bibinfo {author} {\bibfnamefont {Alfred~J.}\
  \bibnamefont {Lotka}},\ }\bibfield  {title} {\enquote {\bibinfo {title}
  {Analytical note on certain rhythmic relations in organic systems},}\ }\href
  {http://www.jstor.org/stable/84156} {\bibfield  {journal} {\bibinfo
  {journal} {PNAS}\ }\textbf {\bibinfo {volume} {6}},\ \bibinfo {pages}
  {410--415} (\bibinfo {year} {1920})}\BibitemShut {NoStop}%
\bibitem [{\citenamefont {Lindeman}(1942)}]{Lindeman42}%
  \BibitemOpen
  \bibfield  {author} {\bibinfo {author} {\bibfnamefont {Raymond}\ \bibnamefont
  {Lindeman}},\ }\bibfield  {title} {\enquote {\bibinfo {title} {The
  trophic-dynamic aspect of ecology},}\ }\href
  {http://search.proquest.com/docview/1296353965/} {\bibfield  {journal}
  {\bibinfo  {journal} {Ecology}\ }\textbf {\bibinfo {volume} {23}} (\bibinfo
  {year} {1942})}\BibitemShut {NoStop}%
\bibitem [{\citenamefont {Trebilco}\ \emph {et~al.}(2013)\citenamefont
  {Trebilco}, \citenamefont {Baum}, \citenamefont {Salomon},\ and\
  \citenamefont {Dulvy}}]{Trebilcoetal13}%
  \BibitemOpen
  \bibfield  {author} {\bibinfo {author} {\bibfnamefont {R.}~\bibnamefont
  {Trebilco}}, \bibinfo {author} {\bibfnamefont {J.~K.}\ \bibnamefont {Baum}},
  \bibinfo {author} {\bibfnamefont {A.~K.}\ \bibnamefont {Salomon}}, \ and\
  \bibinfo {author} {\bibfnamefont {N.~K.}\ \bibnamefont {Dulvy}},\ }\bibfield
  {title} {\enquote {\bibinfo {title} {Ecosystem ecology: size-based
  constraints on the pyramids of life},}\ }\href@noop {} {\bibfield  {journal}
  {\bibinfo  {journal} {Trends In Ecology \& Evolution}\ }\textbf {\bibinfo
  {volume} {28}},\ \bibinfo {pages} {423--431} (\bibinfo {year}
  {2013})}\BibitemShut {NoStop}%
\bibitem [{\citenamefont {Colinvaux}\ and\ \citenamefont
  {Barnett}(1979)}]{Colinvaux79}%
  \BibitemOpen
  \bibfield  {author} {\bibinfo {author} {\bibfnamefont {P.~A.}\ \bibnamefont
  {Colinvaux}}\ and\ \bibinfo {author} {\bibfnamefont {B.~D.}\ \bibnamefont
  {Barnett}},\ }\bibfield  {title} {\enquote {\bibinfo {title} {Lindeman and
  the ecological efficiency of wolves},}\ }\href@noop {} {\bibfield  {journal}
  {\bibinfo  {journal} {The American Naturalist}\ }\textbf {\bibinfo {volume}
  {114}},\ \bibinfo {pages} {707--718} (\bibinfo {year} {1979})}\BibitemShut
  {NoStop}%
\bibitem [{\citenamefont {Hanski}\ \emph {et~al.}(2001)\citenamefont {Hanski},
  \citenamefont {Henttonen}, \citenamefont {Korpim\"aki}, \citenamefont
  {Oksanen},\ and\ \citenamefont {Turchin}}]{Hanskietal01}%
  \BibitemOpen
  \bibfield  {author} {\bibinfo {author} {\bibfnamefont {Ilkka}\ \bibnamefont
  {Hanski}}, \bibinfo {author} {\bibfnamefont {Heikki}\ \bibnamefont
  {Henttonen}}, \bibinfo {author} {\bibfnamefont {Erkki}\ \bibnamefont
  {Korpim\"aki}}, \bibinfo {author} {\bibfnamefont {Lauri}\ \bibnamefont
  {Oksanen}}, \ and\ \bibinfo {author} {\bibfnamefont {Peter}\ \bibnamefont
  {Turchin}},\ }\bibfield  {title} {\enquote {\bibinfo {title} {Small‐rodent
  dynamics and predation},}\ }\href {\doibase 10.2307/2679796} {\bibfield
  {journal} {\bibinfo  {journal} {Ecology}\ }\textbf {\bibinfo {volume} {82}},\
  \bibinfo {pages} {1505--1520} (\bibinfo {year} {2001})}\BibitemShut {NoStop}%
\bibitem [{\citenamefont {Krebs}\ \emph {et~al.}(1995)\citenamefont {Krebs},
  \citenamefont {Boutin}, \citenamefont {Boonstra}, \citenamefont {Sinclair},
  \citenamefont {Smith}, \citenamefont {Dale}, \citenamefont {Martin},\ and\
  \citenamefont {Turkington}}]{Krebsetal95}%
  \BibitemOpen
  \bibfield  {author} {\bibinfo {author} {\bibfnamefont {C.~J.}\ \bibnamefont
  {Krebs}}, \bibinfo {author} {\bibfnamefont {S.}~\bibnamefont {Boutin}},
  \bibinfo {author} {\bibfnamefont {R.}~\bibnamefont {Boonstra}}, \bibinfo
  {author} {\bibfnamefont {A.~R.}\ \bibnamefont {Sinclair}}, \bibinfo {author}
  {\bibfnamefont {J.~N.}\ \bibnamefont {Smith}}, \bibinfo {author}
  {\bibfnamefont {M.~R.}\ \bibnamefont {Dale}}, \bibinfo {author}
  {\bibfnamefont {K.}~\bibnamefont {Martin}}, \ and\ \bibinfo {author}
  {\bibfnamefont {R.}~\bibnamefont {Turkington}},\ }\bibfield  {title}
  {\enquote {\bibinfo {title} {Impact of food and predation on the snowshoe
  hare cycle},}\ }\href@noop {} {\bibfield  {journal} {\bibinfo  {journal}
  {Science (New York, N.Y.)}\ }\textbf {\bibinfo {volume} {269}} (\bibinfo
  {year} {1995})}\BibitemShut {NoStop}%
\bibitem [{\citenamefont {Rinaldi}\ and\ \citenamefont
  {Muratori}(1992)}]{RinaldiMuratori92}%
  \BibitemOpen
  \bibfield  {author} {\bibinfo {author} {\bibfnamefont {S.}~\bibnamefont
  {Rinaldi}}\ and\ \bibinfo {author} {\bibfnamefont {S.}~\bibnamefont
  {Muratori}},\ }\bibfield  {title} {\enquote {\bibinfo {title} {Slow-fast
  limit cycles in predator-prey models},}\ }\href@noop {} {\bibfield  {journal}
  {\bibinfo  {journal} {Ecological Modelling}\ }\textbf {\bibinfo {volume}
  {61}},\ \bibinfo {pages} {287--308} (\bibinfo {year} {1992})}\BibitemShut
  {NoStop}%
\bibitem [{\citenamefont {Sunquist}(2002)}]{SunquistSunquist02}%
  \BibitemOpen
  \bibfield  {author} {\bibinfo {author} {\bibfnamefont {Melvin~E}\
  \bibnamefont {Sunquist}},\ }\href@noop {} {\emph {\bibinfo {title} {Wild cats
  of the world}}}\ (\bibinfo  {publisher} {University of Chicago Press},\
  \bibinfo {address} {Chicago, Ill},\ \bibinfo {year} {2002})\BibitemShut
  {NoStop}%
\bibitem [{\citenamefont {Turchin}\ \emph {et~al.}(2000)\citenamefont
  {Turchin}, \citenamefont {Oksanen}, \citenamefont {Ekerholm}, \citenamefont
  {Oksanen},\ and\ \citenamefont {Henttonen}}]{Turchin00}%
  \BibitemOpen
  \bibfield  {author} {\bibinfo {author} {\bibfnamefont {P.}~\bibnamefont
  {Turchin}}, \bibinfo {author} {\bibfnamefont {L.}~\bibnamefont {Oksanen}},
  \bibinfo {author} {\bibfnamefont {P.}~\bibnamefont {Ekerholm}}, \bibinfo
  {author} {\bibfnamefont {T.}~\bibnamefont {Oksanen}}, \ and\ \bibinfo
  {author} {\bibfnamefont {H.}~\bibnamefont {Henttonen}},\ }\bibfield  {title}
  {\enquote {\bibinfo {title} {Are lemmings prey or predators?}}\ }\href@noop
  {} {\bibfield  {journal} {\bibinfo  {journal} {Nature}\ }\textbf {\bibinfo
  {volume} {405}} (\bibinfo {year} {2000})}\BibitemShut {NoStop}%
\bibitem [{\citenamefont {Hanski}\ \emph {et~al.}(1991)\citenamefont {Hanski},
  \citenamefont {Hansson},\ and\ \citenamefont {Henttonen}}]{Hanski91}%
  \BibitemOpen
  \bibfield  {author} {\bibinfo {author} {\bibfnamefont {I.}~\bibnamefont
  {Hanski}}, \bibinfo {author} {\bibfnamefont {L.}~\bibnamefont {Hansson}}, \
  and\ \bibinfo {author} {\bibfnamefont {H.}~\bibnamefont {Henttonen}},\
  }\bibfield  {title} {\enquote {\bibinfo {title} {Specialist predators,
  generalist predators, and the microtine rodent cycle},}\ }\href
  {http://search.proquest.com/docview/1300205319/} {\bibfield  {journal}
  {\bibinfo  {journal} {The Journal of Animal Ecology}\ }\textbf {\bibinfo
  {volume} {60}} (\bibinfo {year} {1991})}\BibitemShut {NoStop}%
\bibitem [{\citenamefont {Gilg}\ \emph {et~al.}(2003)\citenamefont {Gilg},
  \citenamefont {Hanski},\ and\ \citenamefont {Sittler}}]{Gilg03}%
  \BibitemOpen
  \bibfield  {author} {\bibinfo {author} {\bibfnamefont {Olivier}\ \bibnamefont
  {Gilg}}, \bibinfo {author} {\bibfnamefont {Ilkka}\ \bibnamefont {Hanski}}, \
  and\ \bibinfo {author} {\bibfnamefont {Beno\^it}\ \bibnamefont {Sittler}},\
  }\bibfield  {title} {\enquote {\bibinfo {title} {Cyclic dynamics in a simple
  vertebrate predator-prey community},}\ }\href@noop {} {\bibfield  {journal}
  {\bibinfo  {journal} {Science (New York, N.Y.)}\ }\textbf {\bibinfo {volume}
  {302}} (\bibinfo {year} {2003})}\BibitemShut {NoStop}%
\bibitem [{\citenamefont {Henttonen}\ \emph {et~al.}(1987)\citenamefont
  {Henttonen}, \citenamefont {Oksanen}, \citenamefont {Jortikka},\ and\
  \citenamefont {Haukisalmi}}]{Henttonen87}%
  \BibitemOpen
  \bibfield  {author} {\bibinfo {author} {\bibfnamefont {Heikki}\ \bibnamefont
  {Henttonen}}, \bibinfo {author} {\bibfnamefont {Tarja}\ \bibnamefont
  {Oksanen}}, \bibinfo {author} {\bibfnamefont {Aarre}\ \bibnamefont
  {Jortikka}}, \ and\ \bibinfo {author} {\bibfnamefont {Voitto}\ \bibnamefont
  {Haukisalmi}},\ }\bibfield  {title} {\enquote {\bibinfo {title} {How much do
  weasels shape microtine cycles in the northern fennoscandian taiga?}}\
  }\href@noop {} {\bibfield  {journal} {\bibinfo  {journal} {Oikos}\ }\textbf
  {\bibinfo {volume} {50}},\ \bibinfo {pages} {353--365} (\bibinfo {year}
  {1987})}\BibitemShut {NoStop}%
\bibitem [{\citenamefont {Carlsen}\ \emph {et~al.}(1999)\citenamefont
  {Carlsen}, \citenamefont {Lodal}, \citenamefont {Leirs},\ and\ \citenamefont
  {Jensen}}]{Carlsen99}%
  \BibitemOpen
  \bibfield  {author} {\bibinfo {author} {\bibfnamefont {Michael}\ \bibnamefont
  {Carlsen}}, \bibinfo {author} {\bibfnamefont {Jens}\ \bibnamefont {Lodal}},
  \bibinfo {author} {\bibfnamefont {Herwig}\ \bibnamefont {Leirs}}, \ and\
  \bibinfo {author} {\bibfnamefont {Thomas~Secher}\ \bibnamefont {Jensen}},\
  }\bibfield  {title} {\enquote {\bibinfo {title} {The effect of predation risk
  on body weight in the field vole, microtus agrestis},}\ }\href@noop {}
  {\bibfield  {journal} {\bibinfo  {journal} {Oikos}\ }\textbf {\bibinfo
  {volume} {87}},\ \bibinfo {pages} {277--285} (\bibinfo {year}
  {1999})}\BibitemShut {NoStop}%
\bibitem [{\citenamefont {Moen}\ \emph {et~al.}(2010)\citenamefont {Moen},
  \citenamefont {Rasmussen}, \citenamefont {Burdett},\ and\ \citenamefont
  {Pelican}}]{Moen10}%
  \BibitemOpen
  \bibfield  {author} {\bibinfo {author} {\bibfnamefont {Ron}\ \bibnamefont
  {Moen}}, \bibinfo {author} {\bibfnamefont {James~M.}\ \bibnamefont
  {Rasmussen}}, \bibinfo {author} {\bibfnamefont {Christopher~L.}\ \bibnamefont
  {Burdett}}, \ and\ \bibinfo {author} {\bibfnamefont {Katharine~M.}\
  \bibnamefont {Pelican}},\ }\bibfield  {title} {\enquote {\bibinfo {title}
  {Hematology, serum chemistry, and body mass of free-ranging and captive
  canada lynx in minnesota},}\ }\href@noop {} {\bibfield  {journal} {\bibinfo
  {journal} {Journal of Wildlife Diseases}\ }\textbf {\bibinfo {volume} {46}},\
  \bibinfo {pages} {13--22} (\bibinfo {year} {2010})}\BibitemShut {NoStop}%
\bibitem [{\citenamefont {Gillingham}(1984)}]{Gillingham84}%
  \BibitemOpen
  \bibfield  {author} {\bibinfo {author} {\bibfnamefont {Bruce~J.}\
  \bibnamefont {Gillingham}},\ }\bibfield  {title} {\enquote {\bibinfo {title}
  {Meal size and feeding rate in the least weasel (mustela nivalis)},}\
  }\href@noop {} {\bibfield  {journal} {\bibinfo  {journal} {Journal of
  Mammalogy}\ }\textbf {\bibinfo {volume} {65}},\ \bibinfo {pages} {517--519}
  (\bibinfo {year} {1984})}\BibitemShut {NoStop}%
\bibitem [{\citenamefont {Seal}\ and\ \citenamefont {Mech}(1983)}]{Seal83}%
  \BibitemOpen
  \bibfield  {author} {\bibinfo {author} {\bibfnamefont {U.~S.}\ \bibnamefont
  {Seal}}\ and\ \bibinfo {author} {\bibfnamefont {L.~D.}\ \bibnamefont
  {Mech}},\ }\bibfield  {title} {\enquote {\bibinfo {title} {Blood indicators
  of seasonal metabolic patterns in captive adult gray wolves},}\ }\href@noop
  {} {\bibfield  {journal} {\bibinfo  {journal} {The Journal of Wildlife
  Management}\ }\textbf {\bibinfo {volume} {47}},\ \bibinfo {pages} {704--715}
  (\bibinfo {year} {1983})}\BibitemShut {NoStop}%
\bibitem [{\citenamefont {Krebs}\ \emph {et~al.}(2001)\citenamefont {Krebs},
  \citenamefont {Boonstra}, \citenamefont {Boutin},\ and\ \citenamefont
  {Sinclair}}]{Krebsetal01}%
  \BibitemOpen
  \bibfield  {author} {\bibinfo {author} {\bibfnamefont {Charles~J.}\
  \bibnamefont {Krebs}}, \bibinfo {author} {\bibfnamefont {Rudy}\ \bibnamefont
  {Boonstra}}, \bibinfo {author} {\bibfnamefont {Stan}\ \bibnamefont {Boutin}},
  \ and\ \bibinfo {author} {\bibfnamefont {A.R.E.}\ \bibnamefont {Sinclair}},\
  }\bibfield  {title} {\enquote {\bibinfo {title} {What drives the 10-year
  cycle of snowshoe hares?}}\ }\href@noop {} {\bibfield  {journal} {\bibinfo
  {journal} {BioScience}\ }\textbf {\bibinfo {volume} {51}},\ \bibinfo {pages}
  {25--35} (\bibinfo {year} {2001})}\BibitemShut {NoStop}%
\bibitem [{\citenamefont {Krebs}(2011)}]{Krebs11}%
  \BibitemOpen
  \bibfield  {author} {\bibinfo {author} {\bibfnamefont {Charles~J.}\
  \bibnamefont {Krebs}},\ }\bibfield  {title} {\enquote {\bibinfo {title} {Of
  lemmings and snowshoe hares: the ecology of northern canada},}\ }\href@noop
  {} {\bibfield  {journal} {\bibinfo  {journal} {Proceedings of the Royal
  Society B}\ }\textbf {\bibinfo {volume} {278}},\ \bibinfo {pages} {481--489}
  (\bibinfo {year} {2011})}\BibitemShut {NoStop}%
\bibitem [{\citenamefont {Mowat}\ \emph {et~al.}(2000)\citenamefont {Mowat},
  \citenamefont {O~'~Donoghue},\ and\ \citenamefont {Poole}}]{Mowatetal00}%
  \BibitemOpen
  \bibfield  {author} {\bibinfo {author} {\bibfnamefont {Garth}\ \bibnamefont
  {Mowat}}, \bibinfo {author} {\bibfnamefont {Mark}\ \bibnamefont
  {O~'~Donoghue}}, \ and\ \bibinfo {author} {\bibfnamefont {Kim}\ \bibnamefont
  {Poole}},\ }\bibfield  {title} {\enquote {\bibinfo {title} {Ecology of lynx
  in northern canada and alaska},}\ }in\ \href@noop {} {\emph {\bibinfo
  {booktitle} {Ecology and conservation of lynx in the United States}}},\
  \bibinfo {editor} {edited by\ \bibinfo {editor} {\bibfnamefont {L.F.}\
  \bibnamefont {Ruggiero}}, \bibinfo {editor} {\bibfnamefont {K.B.}\
  \bibnamefont {Aubry}}, \bibinfo {editor} {\bibfnamefont {S.W.}\ \bibnamefont
  {Buskirk}}, \bibinfo {editor} {\bibfnamefont {G.~M.}\ \bibnamefont
  {Koehler}}, \bibinfo {editor} {\bibfnamefont {C.~J.}\ \bibnamefont {Krebs}},
  \bibinfo {editor} {\bibfnamefont {K.S.}\ \bibnamefont {McKelvey}}, \ and\
  \bibinfo {editor} {\bibfnamefont {J.R.}\ \bibnamefont {Squires}}}\ (\bibinfo
  {publisher} {Boulder: University of Colorado Press},\ \bibinfo {year}
  {2000})\ Chap.~\bibinfo {chapter} {9}, pp.\ \bibinfo {pages}
  {265--306}\BibitemShut {NoStop}%
\bibitem [{\citenamefont {Smith}\ \emph {et~al.}(1980)\citenamefont {Smith},
  \citenamefont {Hubartt},\ and\ \citenamefont {Shoemaker}}]{Smithetal80}%
  \BibitemOpen
  \bibfield  {author} {\bibinfo {author} {\bibfnamefont {Ronald~L.}\
  \bibnamefont {Smith}}, \bibinfo {author} {\bibfnamefont {Dennis~J.}\
  \bibnamefont {Hubartt}}, \ and\ \bibinfo {author} {\bibfnamefont
  {Russell~L.}\ \bibnamefont {Shoemaker}},\ }\bibfield  {title} {\enquote
  {\bibinfo {title} {Seasonal changes in weight, cecal length, and pancreatic
  function of snowshoe hares},}\ }\href@noop {} {\bibfield  {journal} {\bibinfo
   {journal} {The Journal of Wildlife Management}\ }\textbf {\bibinfo {volume}
  {44}},\ \bibinfo {pages} {719--724} (\bibinfo {year} {1980})}\BibitemShut
  {NoStop}%
\bibitem [{\citenamefont {Vucetich}\ and\ \citenamefont
  {Peterson}(2004)}]{VucetichPeterson04}%
  \BibitemOpen
  \bibfield  {author} {\bibinfo {author} {\bibfnamefont {John~A.}\ \bibnamefont
  {Vucetich}}\ and\ \bibinfo {author} {\bibfnamefont {Rolf~O.}\ \bibnamefont
  {Peterson}},\ }\bibfield  {title} {\enquote {\bibinfo {title} {The influence
  of prey consumption and demographic stochasticity on population growth rate
  of isle royale wolves (canis lupus)},}\ }\href@noop {} {\bibfield  {journal}
  {\bibinfo  {journal} {Oikos}\ }\textbf {\bibinfo {volume} {107}},\ \bibinfo
  {pages} {309--320} (\bibinfo {year} {2004})}\BibitemShut {NoStop}%
\bibitem [{\citenamefont {Cable}\ \emph {et~al.}(2007)\citenamefont {Cable},
  \citenamefont {Enquist},\ and\ \citenamefont {Moses}}]{Cableetal07}%
  \BibitemOpen
  \bibfield  {author} {\bibinfo {author} {\bibfnamefont {Jessica~M.}\
  \bibnamefont {Cable}}, \bibinfo {author} {\bibfnamefont {Brian~J.}\
  \bibnamefont {Enquist}}, \ and\ \bibinfo {author} {\bibfnamefont
  {Melanie~E.}\ \bibnamefont {Moses}},\ }\bibfield  {title} {\enquote {\bibinfo
  {title} {The allometry of host-pathogen interactions (allometry and
  disease)},}\ }\href@noop {} {\bibfield  {journal} {\bibinfo  {journal} {PLoS
  ONE}\ }\textbf {\bibinfo {volume} {2}} (\bibinfo {year} {2007})}\BibitemShut
  {NoStop}%
\bibitem [{\citenamefont {Dobson}(2004)}]{Dobson04}%
  \BibitemOpen
  \bibfield  {author} {\bibinfo {author} {\bibfnamefont {A.}~\bibnamefont
  {Dobson}},\ }\bibfield  {title} {\enquote {\bibinfo {title} {Population
  dynamics of pathogens with multiple host species},}\ }\href@noop {}
  {\bibfield  {journal} {\bibinfo  {journal} {The American Naturalist}\
  }\textbf {\bibinfo {volume} {164}},\ \bibinfo {pages} {64--78} (\bibinfo
  {year} {2004})}\BibitemShut {NoStop}%
\bibitem [{\citenamefont {Hsieh}\ and\ \citenamefont
  {Hsiao}(2008)}]{HsiehHsiao08}%
  \BibitemOpen
  \bibfield  {author} {\bibinfo {author} {\bibfnamefont {Ying-Hen}\
  \bibnamefont {Hsieh}}\ and\ \bibinfo {author} {\bibfnamefont {Chin-Kuei}\
  \bibnamefont {Hsiao}},\ }\bibfield  {title} {\enquote {\bibinfo {title}
  {Predator-prey model with disease infection in both populations},}\
  }\href@noop {} {\bibfield  {journal} {\bibinfo  {journal} {Mathematical
  Medicine and Biology: A Journal of the IMA}\ }\textbf {\bibinfo {volume}
  {25}},\ \bibinfo {pages} {247--266} (\bibinfo {year} {2008})}\BibitemShut
  {NoStop}%
\end{thebibliography}
\end{document}